\documentclass[vci-paper]{vciart}
\usepackage{amssymb}
\usepackage{epsfig}

\begin{document}
\begin{frontmatter}
\title{The HADES Tracking System}
\author[E]{C.~M{\"u}ntz$^{*,}$}, \author[E]{J.~Markert},
\author[C]{G.~Agakichiev},
\author[F]{H.~Alvarez-Pol}, \author[A]{E.~Badura}, \author[A]{J.~Bielcik},
\author[A]{H.~Bokemeyer}, \author[B]{J.-L.~Boyard}, \author[C]{V.~Chepurnov},
\author[C]{S.~Chernenko}, \author[A]{H.~Daues}, \author[D]{F.~Dohrmann}, \author[D]{W.~Enghardt},
\author[C]{O.~Fateev}, \author[A]{C.~Garabatos}, \author[C]{L.~Glonti}, \author[D]{E.~Grosse},
\author[A]{J.~Hehner}, \author[D]{K.~Heidel}, \author[B]{T.~Hennino}, \author[A]{J.~Hoffmann},
\author[C]{A.~Ierusalimov}, \author[D]{B.~K{\"a}mpfer}, \author[D]{K.~Kanaki},
\author[A]{W.~Koenig}, \author[D]{R.~Kotte},
\author[D]{L.~Naumann},
\author[A]{W.~Ott}, \author[E]{Y.C.~Pachmayer}, \author[C]{V.~Pechenov},
\author[B]{J.~Pouthas}, \author[B]{B.~Ramstein},
\author[E]{K.~Rosenkranz},
\author[B]{P.~Rosier}, \author[B]{M.~Roy-Stephan}, \author[A]{A.~Rustamov},
\author[D]{A.~Sadovski}, \author[C]{L.~Smykov}, \author[D]{M.~Sobiella},
\author[A]{H.~Stelzer}, \author[E]{H.~Stroebele},\author[A]{J.~Stroth},
\author[A]{C.~Sturm}, \author[A]{M.~Sudol},
\author[E]{J.~W{\"u}stenfeld}, \author[C]{Y.~Zanevsky},\author[A]{P.~Zumbruch}
\address[A]{Gesellschaft f{\"u}r Schwerionenforschung, 64291 Darmstadt, Germany}
\address[B]{IPN, 91406 Orsay, France}
\address[C]{Joint Institut of Nuclear Research, 141980 Dubna, Russia}
\address[D]{IKH, Forschungszentrum Rossendorf, 01314 Dresden, Germany}
\address[E]{Johann Wolfgang Goethe-Universit{\"a}t, 60486 Frankfurt, Germany}
\address[F]{Universidade de Santiago de Compostela, 15706 Santiago de Compostela, Spain}
\small{$^*$corresponding author, email: c.muentz@gsi.de}
\begin{abstract}
The HADES dielectron spectrometer has recently launched its
physics program at the heavy ion synchrotron SIS at GSI Darmstadt.
The spectroscopy of vector mesons in heavy ion collisions via
their dielectron decay channel makes great demands on the HADES
tracking system regarding acceptance and spatial resolution. The
tracking system is formed out of 24 low-mass, trapezoidal
multi-layer drift chambers providing about 30~m$^2$ of active
area. Low multiple scattering in the in total four planes of drift
chambers before and after the magnetic field is ensured by using
helium-based gas mixtures and aluminum cathode and field wires.
First in-beam performance results are contrasted with expectations
from simulations. Emphasis is placed on the energy loss
information, exploring its relevance regarding track recognition.
\end{abstract}
\end{frontmatter}
\small
\section{Introduction}
The HADES dielectron spectrometer~\cite{hades} at the heavy ion
synchrotron SIS at GSI Darmstadt has taken up its physics program
in several beam time campaigns, starting late 2001, focusing on
C+C reactions at 1 and 2~AGeV incident energy. Recently, a beam
time studying p+p reactions in the same energy regime
has been successfully conducted. \\
The spectroscopy of vector mesons in heavy ion collisions via
their dielectron decay channel defines the decisive design and
performance constraints on the HADES tracking system. An intrinsic
spatial cell resolution of better than 140~$\mu$m along with the
reduction of multiple scattering in detector materials and gas,
high efficiency and a large acceptance are crucial requirements
for the success of the experimental program. Extended design
studies and prototyping~\cite{mdc1,mdc2} preceded the production
of the 24 drift chambers of four different sizes, conducted by GSI
Darmstadt, LHE/JINR Dubna, FZ Rossendorf~\cite{mdc3},
IPN Orsay and University of Frankfurt.\\
Presently 22 chambers are in-place and do already now allow for
high-precision tracking in two-third of the full azimuth. In the
following the tracking system and its design parameters are
introduced. Then, first results are briefly reviewed, highlighting
data on the capability of the drift chamber energy loss signal
regarding track matching, and particle identification in general.
\section{The HADES drift chambers}
\subsection{Detector characteristics}
The HADES tracking system consists of 24 trapezoidal, planar
multi-layer drift chambers (MDC) symmetrically arranged in six
identical sectors. It provides a polar angles coverage between 18
and 85 degrees around the beam axis, forming four tracking planes
(I-IV) of increasing size. In each sector, two chambers (plane I
and II) are located in front of and two (plane III and IV) behind
the toroidal magnetic field of the super-conducting magnet, as
shown in figure~1.
To cope with ambiguities in track reconstruction in the high
multiplicity environment of a heavy ion reaction (for central
Au+Au collisions at 1~AGeV incident energy a maximum cell
occupancy of 30\% is estimated), all chambers are composed of six
sense/field wire layers oriented in five different stereo angles.
This allows for maximum spatial resolution in polar direction,
which points in the direction of the momentum kick. All four
chamber types contain about 1100 drift cells each, with increasing
size from 5x5~mm$^2$ (plane I) to 14x10~mm$^2$ (plane IV). The
chambers provide active areas from 0.35~m$^2$ up to 3.2~m$^2$,
thus covering the same solid angle per sector, respectively. The
main feature of the design and the operation parameters of the
chambers is the consequent implementation of the low-mass concept,
as result of the R\&D phase~\cite{mdc1}. To meet these
requirements (i) cathode and field wires are made of annealed
aluminum (plane I-III: bare, IV: gold-plated) with 80~$\mu$m and
100~$\mu$m diameter, and with initial tensions between 80 and
150~cN, depending on the chamber type. In addition, (ii)
helium-based counting gas (helium:isobutane = 60:40) is used. The
entrance windows are made of 12~$\mu$m aluminized mylar. The
20~$\mu$m (planes I-III) and 30~$\mu$m (plane IV) thick
gold-plated tungsten sense wires are strained with an initial
tension of 40 and 50~cN, respectively. To compensate for the total
wire tension after being released from the assembly table, all
chamber frames have been pre-stressed before wire gluing. Together
with the requirements concerning the acceptance this resulted in a
sophisticated layer frame design with only 3~cm width for the
inner-most chambers. The total detector thickness per chamber in
units of radiation length is below 5x10$^{-4}$ and hence low
multiple scattering guarantees the
momentum resolution needed to accomplish the physics demands.\\
\subsection{Long-term stability}
The experiments with HADES are expected to run at least ten years.
Creeping of the aluminum wires and ageing are the main concerns
with respect to the long-term stability of the chambers. Creeping
has been systematically investigated in tension loss test series,
yielding a 10\% loss in tension within five years. This has been
confirmed by remeasurements in one chamber of
plane~III~\cite{mdc3}. Ageing is mainly caused by the accumulated
dose in combination with the materials used for construction and
operating the chambers. For example, epoxy from Araldite is used
for gluing the wires on the frames from Stesalit. The gas system
is running in a re-flow mode with typically 10~\% fresh gas and
continuous purification. The drift velocity monitors~\cite{CLipp}
provide a very sensitive control of the gas quality by measuring
the drift velocity with a precision of better than 0.2~\%. In
addition, the simultaneous monitoring of the relative gains allows
for conclusions on the gas contamination, e.g. due to oxygen. The
expected maximum dose is in the order of 10~mC per year and
centimeter of the sense wire. An accelerated ageing test with
$^{55}$Fe using two prototype chambers exhibited no
noticeable gain drop ($<$5\%) for a time period of two years of running~\cite{mdc1}.\\
These R\&D results on creeping and ageing, together with the
careful selection of materials and running conditions, suggest
that the projected long-term operation of the HADES tracking
system can be kept.
\subsection{Read-out electronics}
The drift chamber signals are read-out and digitized at the
chamber, not extending into its active area. Hence, special
emphasis was put on the integration of the modular front-end
electronics, realized with analog boards (16 channels) mounted on
digitization boards (64 or 96~channels). Four sense wires are
connected by flexible printed circuits to the analog
boards~\cite{DB}, housing ASD8-B chips~\cite{asd8} (8 channels,
1~fC intrinsic noise, 30~mW/ch, adjustable threshold) for
differential amplification, shaping and discrimination. These
chips deliver logical signals with variable width, equivalent to
the time the shaped signal exceeds the given threshold. The
logical signals are fed to TDC chips (CMOS, 8 channels/chip, 0.5
ns/ch, common-stop, 1 $\mu$s full range, making use of a ring
oscillator) on the digitization boards. This semi-customized ASIC
is multi-hit capable and thus allows to detect also the
time-above-threshold of each hit. Besides spike and zero
suppression this chip offers the possibility of internal
calibration, activated by a separate trigger type. The design of
the front-end electronics was decisively influenced by minimizing
the noise level on-line. In addition, the time-above-threshold
information is a valuable tool to discriminate remaining noise
hits off-line.
\section{Results}
\subsection{Performance}
The chambers have been running in several C+C and p+p experiments
with moderate intensity and occupancy. Layer detection
efficiencies around 98~\% have been deduced. The average spatial
resolutions of a drift cell can be determined with data by means
of a self-tracking method and is between 100 and 130~$\mu$m,
comprising contributions from calibration accuracy and electronic
noise. This is well below the design value of 140~$\mu$m, see
figure~2.
Under clean conditions, with 2.1~GeV/c protons impinging on a
prototype, using a silicon-strip tracker as reference and standard
read-out electronics, intrinsic spatial resolutions around
70~$\mu$m have been reached~\cite{mdc2}, similar to results given
by GARFIELD simulations. Preliminary simulation studies on the
expected track spatial resolution in a segment of two adjacent
chambers yield values below 60~$\mu$m (modules of plane I and II)
in direction of the momentum kick, made possible by the multiple
position measurement per segment, see figure~3. These simulations
have been confirmed by experimental results from tracking,
suggesting an error in measuring the angle of a segment in the
polar direction of about 0.2~mrad.
\subsection{Energy loss and tracking}
Besides the arrival time of the electrons the
time-above-threshold~$\Delta$t of the corresponding signal is
measured for each cell hit. From the latter value the energy loss
can be deduced. $\Delta$t is a measure of the number of produced
primary electron clusters~\cite{jochen}, i.e.~the energy loss of a
particle traversing the drift cell. It is determined by the
ionization density of the particle, and the track topology in a
given cell, which is characterized by the impact angle and the
impact parameter relative to the sense wire. The energy loss along
a track (segment) can be used to resolve ambiguities in matching
track segments from the inner and outer chambers, to support the
particle identification by adding complementary information, and
to reject tracks originating from conversion and Dalitz decays
with small opening angles - the main background in dielectron spectroscopy.\\
To investigate the resolution of the energy loss signal, effects
from the topology of the track as well as of the electron drift
inside a cell have to be corrected for. These topological effects
can be corrected by using the local information on the orientation
and position of the track segment, defined by two adjacent
chambers in one sector, respectively. Here, GARFIELD simulations
on time-to-distance relations in different cell geometries are
employed. Simulation results and measured data coincide to a high
degree~\cite{jochen}. Effects of the electron drift topology on
the energy loss signal have been systematically parameterized
based on data only, and are used for normalization. As a result,
the normalized $\Delta$t can be plotted as a function of the
measured momentum, shown in figure~4.
Clearly, the separation of pions and protons at low momenta can be
seen. Electrons cannot be distinguished from pions. The achieved
resolution ($\sigma$) of the normalized $\Delta$t so far is about
7~\% (protons) and 12~\% (pions), possible improvements are
currently under investigation. However, these results provide
already now confidence in the analysis power of the MDC's energy
loss information, although the chamber design has originally not
been optimized in this respect.
\section{Conclusions}
The low-mass HADES tracking system is operational and the
performance is according to its design values, thus allowing for
high-resolution dielectron spectroscopy. Currently, scenarios are
under debate on the possibility to include the hit information of
the chambers in a higher level trigger. The energy loss signal
measured for each track turned out to be a valuable tool to
improve on resolving ambiguities in tracking and on background
reduction, even though the chamber design has been optimized for
minimizing the detector thickness.

\twocolumn
\section{Figure captions}
\mbox{\epsfig{file=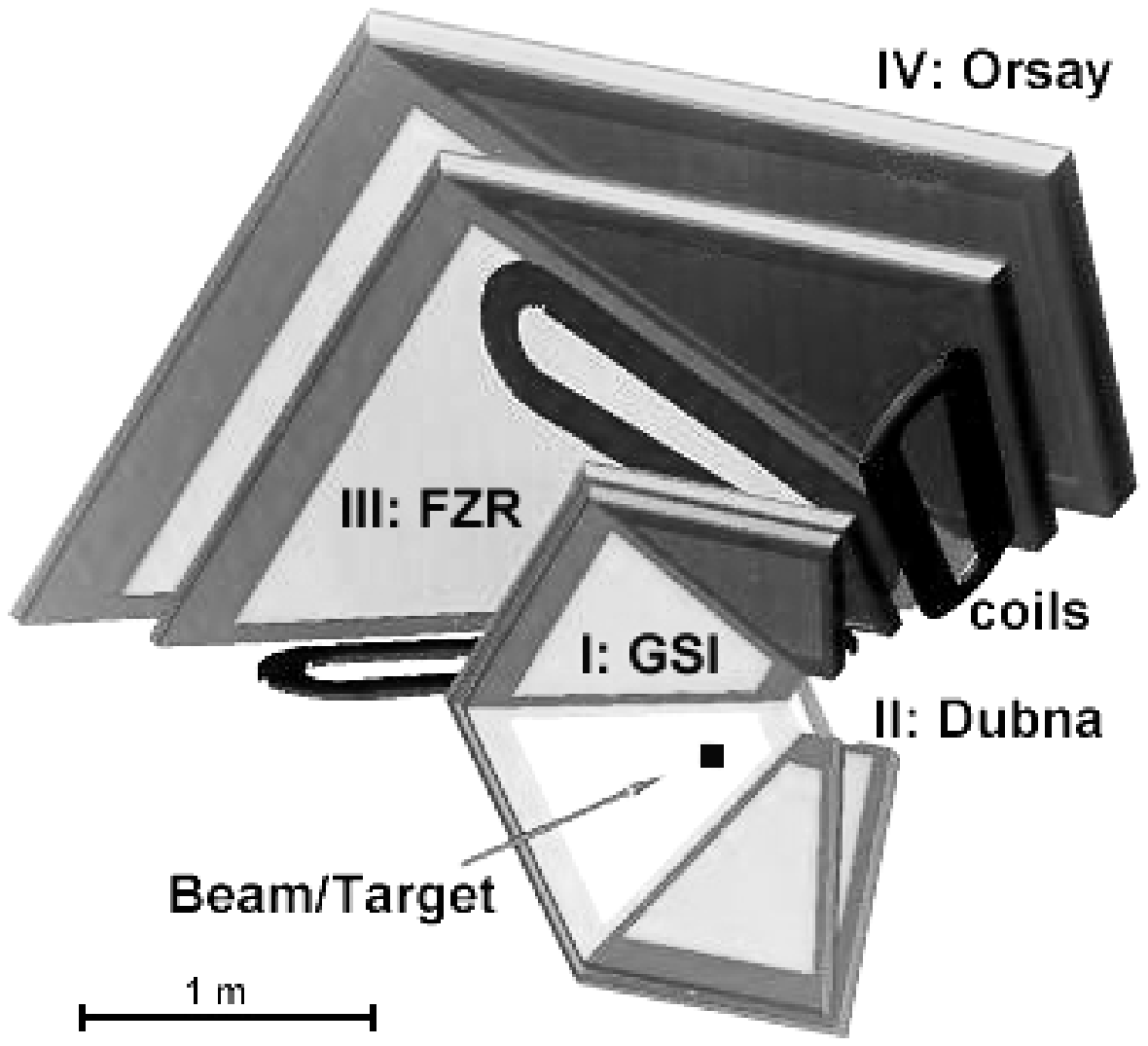,height=5.5cm}}\\
Fig. 1: Schematic view of the HADES tracking system, arranged in a
frustum-like geometry before and behind the magnet coils.
The four chamber construction sites are indicated.\\

\mbox{\epsfig{file=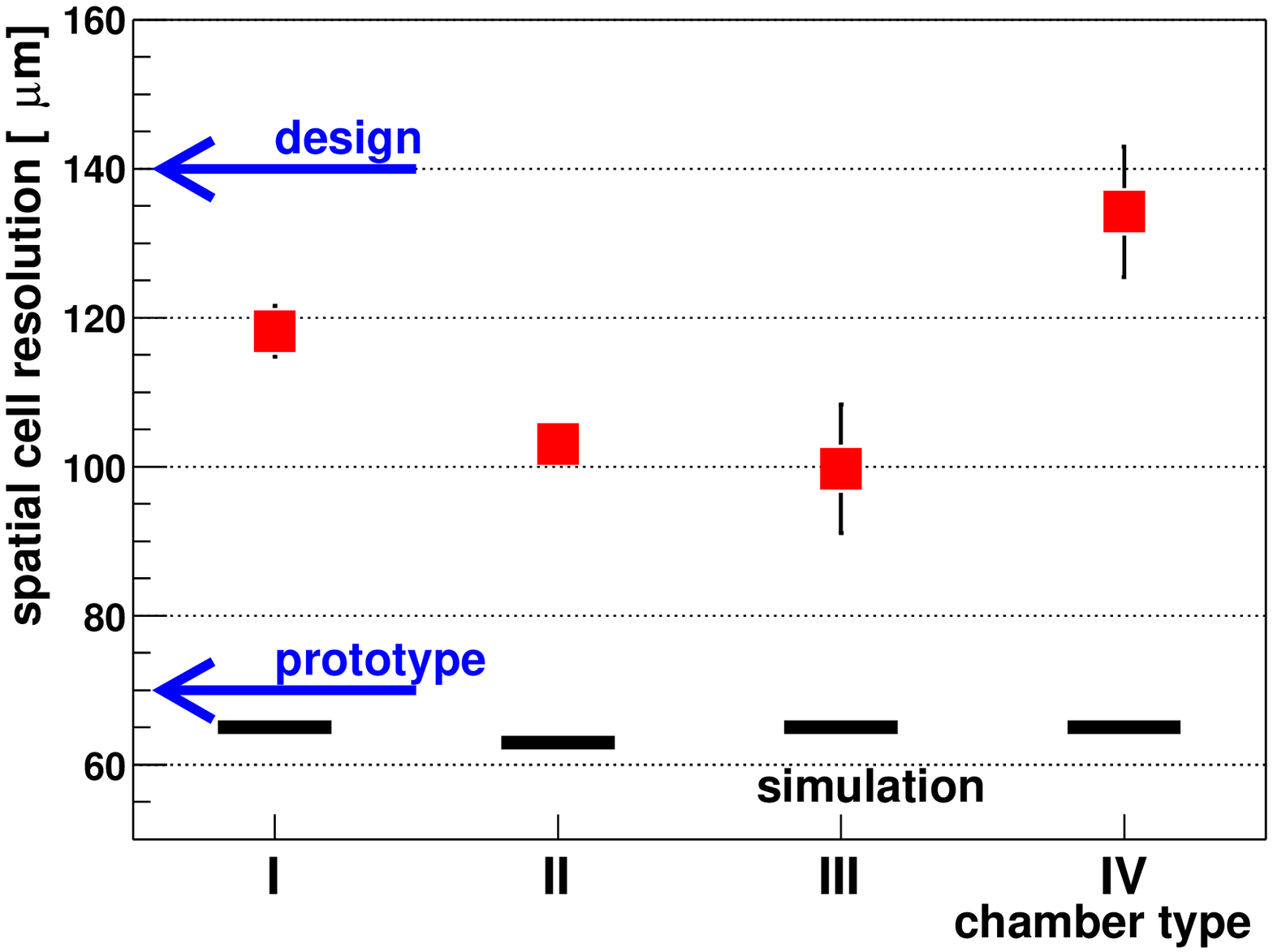,height=4.8cm}}\\
Fig. 2: Average in-beam spatial drift cell resolution (symbols),
for the four chamber types I-IV present in 2001. The error bars
depict the variance of several modules of one type (no systematic
errors, and not corrected for contributions from read-out
electronics and calibration). The data is compared to the design
value, a result from prototype studies, and Garfield simulations.\\

\mbox{\epsfig{file=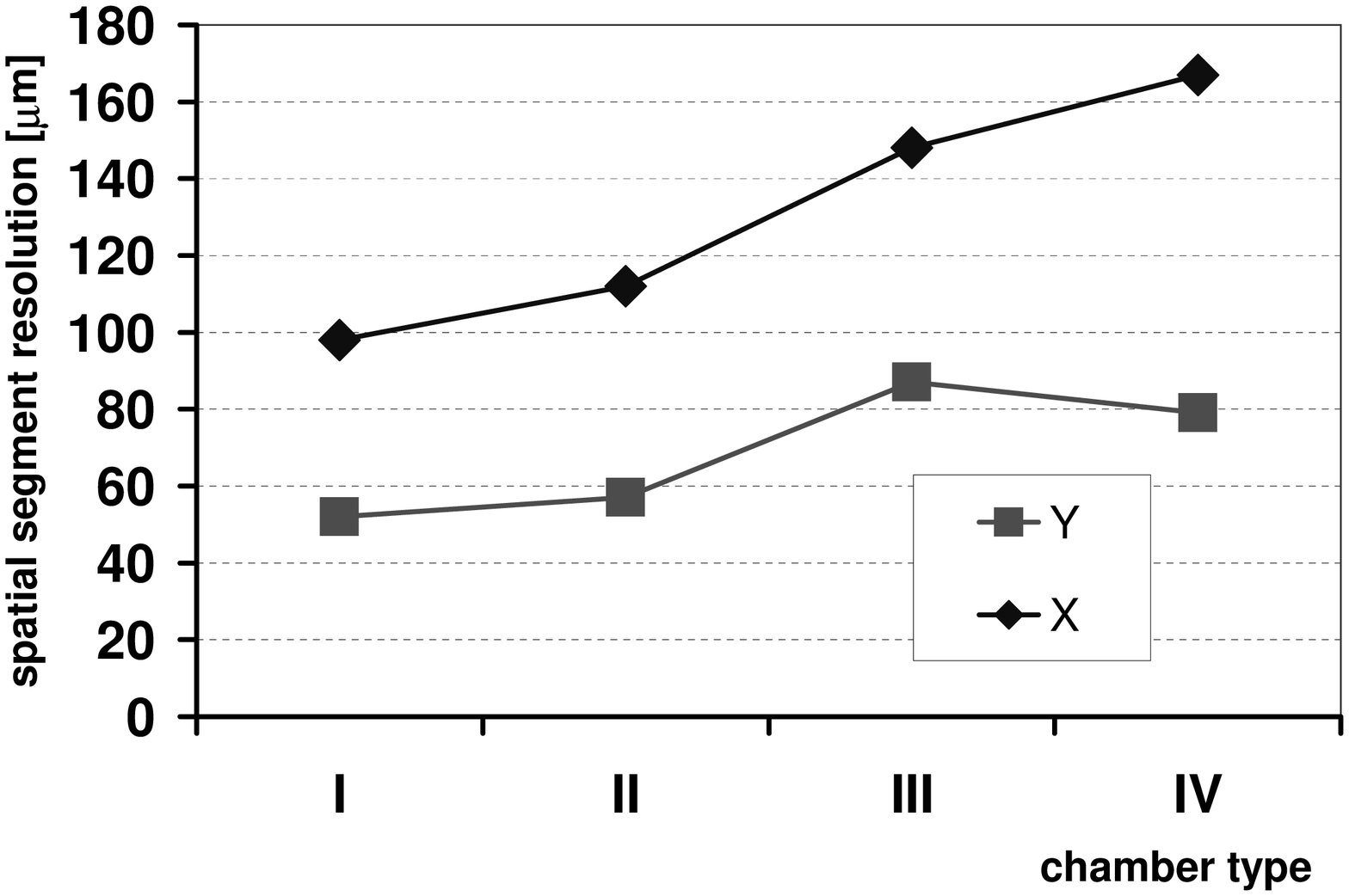,height=4.3cm}}\\
Fig. 3: Preliminary results from simulations: Spatial segment
resolution for the four chamber types (I-IV), in Y (momentum kick
direction) and perpendicular (X).\\

\mbox{\epsfig{file=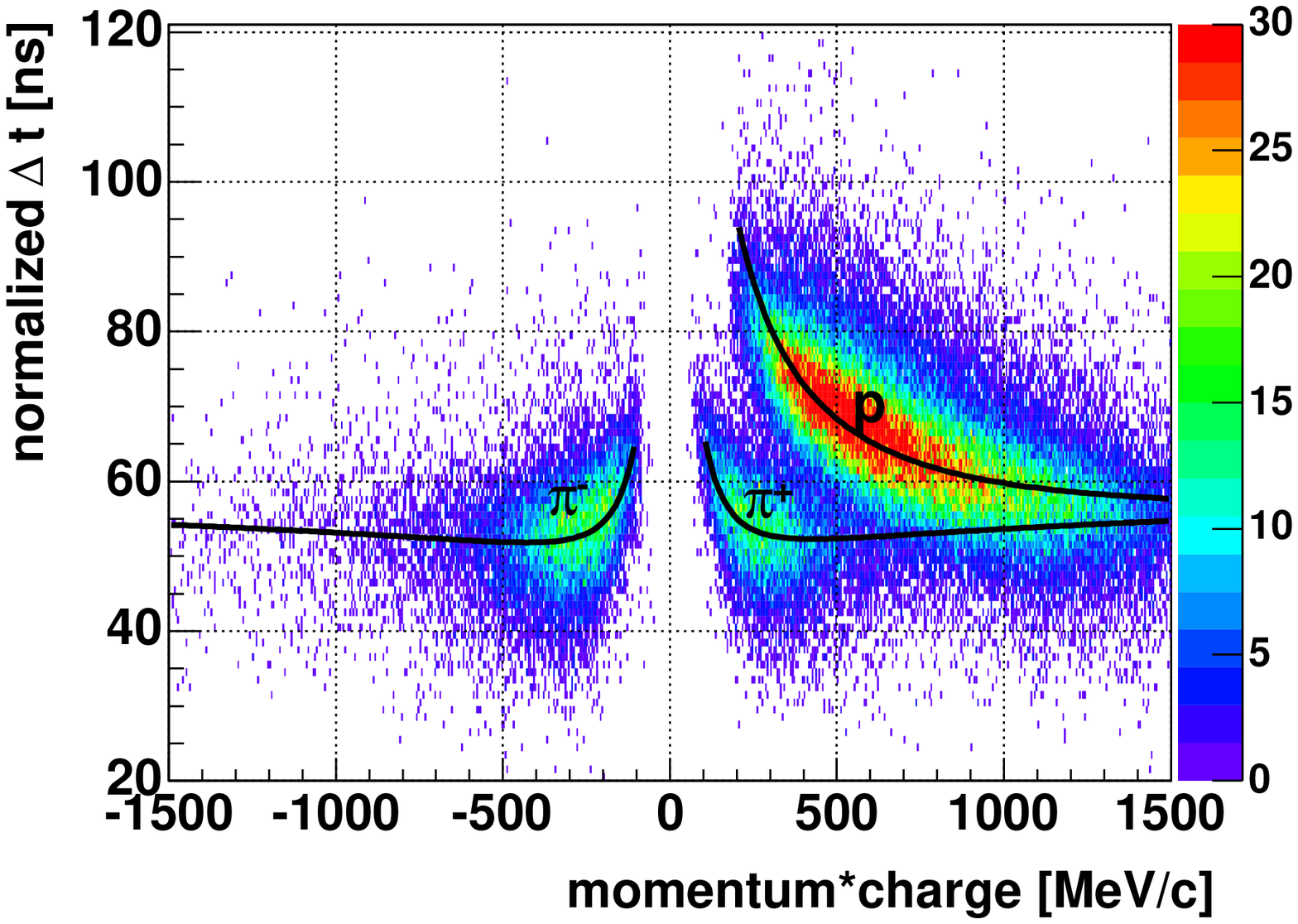,height=4.6cm}}\\
Fig. 4: Data from C+C, 2 AGeV: normalized time-above-threshold
$\Delta$t is plotted as a function of the measured momentum times
the charge for the inner MDC segment (planes I and II). Lines
indicate Bethe-Bloch parameterizations.\\

\end{document}